\documentclass{icrc2009}

\usepackage{graphicx}   
\usepackage[caption=false]{caption}    
\usepackage[font=footnotesize]{subfig} 
\usepackage{fixltx2e}
\usepackage{url}

\newcommand{\shorttitle}[1]%
{\markboth{Proceedings of the 31\MakeLowercase{$^{st}$} ICRC, {\L}\'{o}d\'{z} 2009}{#1} }
\newcommand{\etal}{\MakeLowercase{\textit{et al. }}} 

\begin{document}
\title{Search for Neutrinos from Dark Matter Annihilation in the Sun with the 
Baikal Neutrino Experiment}

\author{\IEEEauthorblockN{A. Avrorin\IEEEauthorrefmark{1},
			  V. Aynutdinov\IEEEauthorrefmark{1},
                          V. Balkanov\IEEEauthorrefmark{1},
                          I. Belolaptikov\IEEEauthorrefmark{4},
			  D. Bogorodsky\IEEEauthorrefmark{2},
                          N. Budnev\IEEEauthorrefmark{2},\\
                          I. Danilchenko\IEEEauthorrefmark{1},
                          G. Domogatsky\IEEEauthorrefmark{1},
			  A. Doroshenko\IEEEauthorrefmark{1},
                          A. Dyachok\IEEEauthorrefmark{2},
			  Zh.-A. Dzhilkibaev\IEEEauthorrefmark{1},\\
                          S. Fialkovsky\IEEEauthorrefmark{6},
			  O. Gaponenko\IEEEauthorrefmark{1},
                          K. Golubkov\IEEEauthorrefmark{4},
                          O. Gress\IEEEauthorrefmark{2},
			  T. Gress\IEEEauthorrefmark{2},
                          O. Grishin\IEEEauthorrefmark{2},\\
			  A. Klabukov\IEEEauthorrefmark{1},
                          A. Klimov\IEEEauthorrefmark{8},
                          A. Kochanov\IEEEauthorrefmark{2},
                          K. Konischev\IEEEauthorrefmark{4},
                          A. Koshechkin\IEEEauthorrefmark{1},
                          V. Kulepov\IEEEauthorrefmark{6},\\
                          D. Kuleshov\IEEEauthorrefmark{1},
                          L. Kuzmichev\IEEEauthorrefmark{3},
                          V. Lyashuk\IEEEauthorrefmark{1},
                          E. Middell\IEEEauthorrefmark{5},
                          S. Mikheyev\IEEEauthorrefmark{1},
                          M. Milenin\IEEEauthorrefmark{6},\\
                          R. Mirgazov\IEEEauthorrefmark{2},
                          E. Osipova\IEEEauthorrefmark{3},
                          G. Pan'kov\IEEEauthorrefmark{2},
                          L. Pan'kov\IEEEauthorrefmark{2},
                          A. Panfilov\IEEEauthorrefmark{1},
                          D. Petuhov\IEEEauthorrefmark{1},\\
                          E. Pliskovsky\IEEEauthorrefmark{4},
                          P. Pokhil\IEEEauthorrefmark{1},
                          V. Poleschuk\IEEEauthorrefmark{1},
                          E. Popova\IEEEauthorrefmark{3},
                          V. Prosin\IEEEauthorrefmark{3},
                          M. Rozanov\IEEEauthorrefmark{7},\\
                          V. Rubtzov\IEEEauthorrefmark{2},
                          A. Sheifler\IEEEauthorrefmark{1},
                          A. Shirokov\IEEEauthorrefmark{3},
                          B. Shoibonov\IEEEauthorrefmark{4},
			  Ch. Spiering\IEEEauthorrefmark{5},
			  O. Suvorova\IEEEauthorrefmark{1},\\
			  B. Tarashansky\IEEEauthorrefmark{2},
			  R. Wischnewski\IEEEauthorrefmark{5},
			  I. Yashin\IEEEauthorrefmark{3},
			  V. Zhukov\IEEEauthorrefmark{1}}
                            \\

\IEEEauthorblockA{\IEEEauthorrefmark{1}Institute for Nuclear Research of Russian Academy of Sciences, Russia}
\IEEEauthorblockA{\IEEEauthorrefmark{2}Irkutsk State University, Irkutsk, Russia}
\IEEEauthorblockA{\IEEEauthorrefmark{3}Skobeltsyn Instutute of Nuclear Physics MSU, Moscow, Russia}
\IEEEauthorblockA{\IEEEauthorrefmark{4}Joint Institute for Nuclear Research, Dubna, Russia}
\IEEEauthorblockA{\IEEEauthorrefmark{5}DESY, Zeuthen, Germany}
\IEEEauthorblockA{\IEEEauthorrefmark{6}Nizhni Novgorod State Technical University, Nizhnij Novgorod, Russia}
\IEEEauthorblockA{\IEEEauthorrefmark{7}St.Petersburg State Marine University, St.Peterburg, Russia}
\IEEEauthorblockA{\IEEEauthorrefmark{8}Kurchatov Institute, Moscow, Russia}}


\shorttitle{A.Avrorin \etal Search for Neutrinos from Dark Matter}
\maketitle

\begin{abstract}
Upward through-going muons in the Lake Baikal Neutrino Experiment arriving from 
the ecliptic plane have been analyzed using NT200 data samples of the years
1998-2002 (1007 live days). We derive upper limits on  
muon fluxes from annihilation processes of hypothetical WIMP dark matter 
particles in the center of the Sun.
\end{abstract}


\section{Introduction}
Since long, galactic rotation curves indicate the existence of non-radiating 
dark matter. This evidence has been hardened also by evidences on larger
cosmic scales, ending up with the largest scales, surveyed with the
help of the cosmic microwave background radiation. In particular, the data
suggest a dominant contribution of non-baryonic, non-relativistic (cold) dark 
matter. Among the favoured DM candidates are weakly interacting massive particles 
(WIMPs), in particular neutralinos as the lightest particles
in many minimal supersymmetric models.

One strategy to search for a neutralino signal is to search for
neutrinos produced in annihilation
processes inside a large gravitational mass like the Sun.
This method is well established since the nineties
using underground and underwater(ice) neutrino telescopes.
The Baikal Neutrino Experiment has published results of a search
for neutrinos from WIMP annihilations in the core of the Earth 
\cite{ref:BaikalDM}, while the
analysis with respect to the Sun is now presented the first time. 
We are looking for high energy neutrinos
from the Sun in excess of the
expected atmospheric neutrinos. The analysis is based on data taken 
with the NT200 neutrino telescope between April 1998 and February 2003.

\section{The Baikal Neutrino Telescope}

Located in Lake Baikal, South-East Siberia, the Neutrino Telescope
NT200 is operated 
underwater
at a depth of \mbox{1.1 km} since 1998. The telescope detects Cherenkov
light from upward and downward-going relativistic muons.
NT200 consists of 192 optical modules (OMs)
arranged pair-wise on 8 strings, with 12 pairs per string. 
The height of the detector
is 72\,m, its diameter is 42\,m. Each OM contains a 37-cm photomultiplier
tube ({\it PMT}). To suppress background from
bioluminescence and dark noise, the two PMTs of a pair are switched in 
coincidence.
Since 2005, the NT200 configuration was upgraded by additional 3 strings, each
100\,m away from the center. This upgraded detector of about 10 Mton 
(named NT200+) serves as a prototype cell for a later Gigaton
volume detector. The status of the Baikal 
Neutrino Telescope is presented at this conference \cite{ref:BaiICRCstatus}.

\section{Selected data samples}
We have analyzed NT200 data collected between April, 1998
and February, 2003, with a total of 1007 live days. Calibration methods
and methods to reconstruct muon tracks have been described elsewhere 
\cite{ref:BaikaL99,ref:BaiNIM06,ref:BaiAP06}.
Our analysis is based on data taken with the {\it muon trigger}. 
It requires \mbox{N$_{hit} \geq n$} within \mbox{500\,ns}, 
where {\it hit} refers
to a pair of OMs coupled in a {\it channel}. Typically \mbox{n} is set to 
\mbox{3 or 4}.
The detector response to atmospheric neutrinos and muons has been
obtained with Monte Carlo simulations based on standard codes like 
{\texttt{CORSIKA }}\cite{corsika} and {\texttt{MUM }}\cite{MUM}, the Bartol 
atmospheric $\nu$ flux \cite{Bartol} and the neutrino cross-sections 
from \cite{Reno96}.

The offline filter which requires at least
6 hits on at least 3 strings ("6/3") selects about
$40\%$ of all triggered events.
To distinguish upward and downward 
going muons on a one-per-million mis-assignment level,
a filter with several levels of quality cuts  
was developed for the atmospheric neutrino ($\nu_{at}$) 
analysis \cite{ref:IgorBel07}.
The atmospheric muons which 
have been mis-reconstructed as upward-going particles are the main source
of background in a search for neutrino induced upward-going muons.
To get the best possible estimator for the 
direction,
we use multiple 
start guesses for the $\chi^{2}$ minimization.
For the final choice of the local minimum of $\chi^{2}$ we use 
quality parameters which are not related to the time information.
At the offline filter level ("6/3") the 
angular resolution ($\Psi$ - r.m.s. mismatch angle) 
is about $14.1^\circ$ for the $\nu_{at}$-sample. 
The present analysis defines two samples - sample A and sample B -
which are optimized for the low and high WIMP-mass region, respectively.
They use further differently tight quality cuts, resulting in 
different background contaminations and sligthly
different angular resolution.
The quality cuts are applied to variables like the 
number of hit channels, the probability of fired channels
to have been hit or not, the actual position of the track
with respect to the detector centre and $\chi^2/d.o.f.$.
To improve the signal-to-background ratio we used only events with
reconstructed zenith angle $\Theta > 100^\circ$. 
This results in 2376 and 510 upward going muons 
for sample A and B, respectively (with
$\nu_{at}$-angular resolutions $\Psi=5.3^\circ$ and $\Psi=3.9^\circ$). 
Both $\Psi$ values are much bigger than the visible size of the Sun. However the angular 
window for a signal search may be even larger, since at least at energies below 100 GeV 
the kinematical angle between neutrino and muon dominates.

\begin{figure}[!t]
  \centering
\includegraphics[width=3.5in]{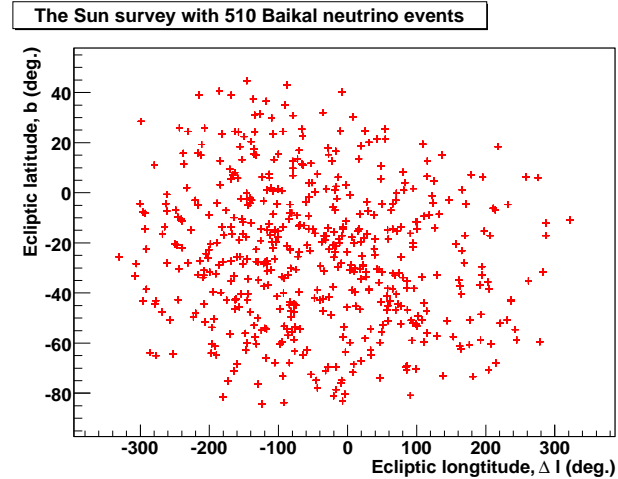}
\caption{Distribution of 510 
upward muon directions in the ecliptic plane, with the Sun
at the coordinate origin.}
\label{surSun510}
\end{figure}

\begin{figure}[!t]
  \centering
\includegraphics[width=3.5in]{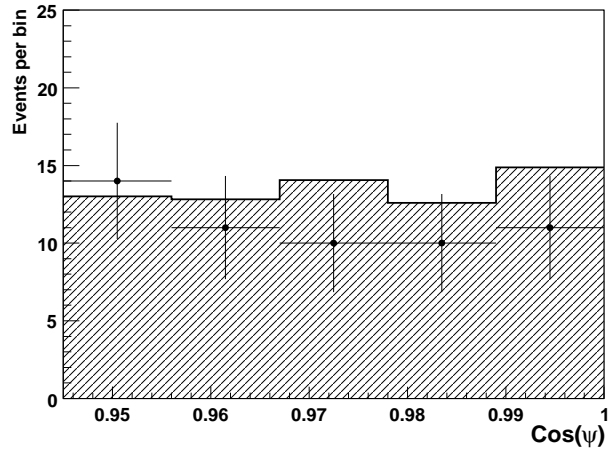}
\caption{Distribution of correlation angles between muons and the Sun
for data sample A of 2376 events: data (bullets) and measured background 
(histogram).}
\label{csnSunMu2376_945}
\end{figure}

\begin{figure}[!t]
  \centering
\includegraphics[width=3.2in]{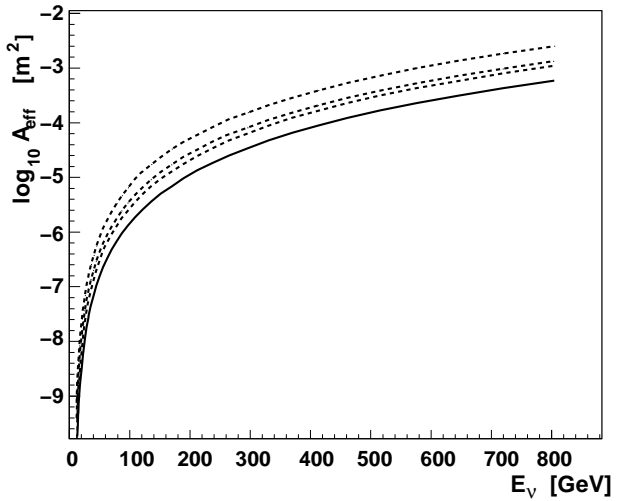}
\caption{Neutrino effective area of NT200
versus neutrino energy for different selection cuts. 
Curves from top:
offline filter level without (short dashed) and with correction (long dashed) 
for the Sun track; 
for event sample B without (dots) and with correction 
(solid line) for the Sun track.}
\label{AreaNeutrinoCuts}
\end{figure}

\section{Sky-plot analysis}
A map of the ecliptic plane centered at the Sun is shown 
in Fig.\ref{surSun510} with 510 upward going
muons. No clustering toward the Sun is observed. 
The distribution of the correlation angles 
between muons and the Sun is shown in Fig.\ref{csnSunMu2376_945}.
The dots refer to angles with the real position of the Sun,
the histogram to angles with 
"fake Suns", defining the expected background behaviour.

No excess is observed, resulting in upper limits on number 
of muons from the Sun.
Table I gives the upper limits at $90\%$ confidence level (c.l.)
for the two selected data samples. We give numbers for three values of
the half cone to the Sun which are used in the dark matter analysis.
To obtain upper limits on the muon numbers we used 
the recipes for low statistics given in
the PDG \cite{ref:PDG08} and well as 
the Feldman-Cousins method
\cite{ref:FeldCou98} (numbers for FC in parentheses).

\begin{table*}[!h]
\begin{center}
\footnotesize\rm
\renewcommand{\tabcolsep}{1.pc} 
\renewcommand{\arraystretch}{1.6} 
\caption{\label{crflux1}  
The $90\%$ c.l. upper limits on the number of muon events
from the Sun for analysis samples A and B, 
calculated according to PDG and (in brackets) FC, respectively.}

\vspace{0.5cm}
\begin{tabular}{|p{5mm}|p{5mm}|p{5mm}|p{7mm}|p{5mm}|p{5mm}|p{7mm}|}
\hline
\multicolumn{1}{|c|}{Half-cone}
&\multicolumn{1}{c}{ }
&\multicolumn{1}{c}{ Sample B }
&\multicolumn{1}{c}{ }
&\multicolumn{1}{c}{ }
&\multicolumn{1}{c}{ Sample A }
&\multicolumn{1}{c|}{ } \\ \hline
{(${\circ}$)}
& {$N_{obs}$} & {$N_{bkg}$}& {$N_{c.l.}$}
& {$N_{obs}$} & {$N_{bkg}$}& {$N_{c.l.}$} \\
\hline
 4 &  1 &  1.5 & 3.1\,(2.9) &  5 &  5.8 &  4.8\,(4.2) \\
15 &  4 &  8.0 & 3.6\,(2.0) & 36 & 47.3 &  7.10 \\
20 & 11 & 16.3 & 4.9\,(2.9) & 78 & 87.7 & 11.50 \\
\hline
\end{tabular}
\end{center}
\end{table*}

\section{Upper limits on a dark matter signal from the Sun}
To derive limits on neutrino and muon fluxes one needs the effective area of the detector. 
Fig.\ref{AreaNeutrinoCuts} shows the NT200 area for neutrinos as a function of neutrino energy.
The expected numbers of muons generated by neutrinos is obtained from the neutrino flux
${\Phi^{\nu}}$ of the considered
source and the effective area $A^{\nu}_{eff}$ of the neutrino telescope:
\begin{equation}
\label{eqn : rate}
N_{\mu}= {\int ^{E_{up}} _{Eth} {A^{\nu}_{eff}(E_{th}, E_{\nu}, \Theta, \phi)} 
\times
{\frac{d^2\Phi^{\nu}}{dE_{\nu}d\Omega} dE_{\nu}d\Omega}}
\end{equation}

For neutrinos originating from WIMP annihilation, the maximum energy
$E_{up}$ is equal to the WIMP mass $m_{WIMP}$, with theory suggesting WIMP
masses in the GeV-TeV range. With increasing
$m_{WIMP}$ more and more channels open up -- both in annihilation
and decay processes. The contribution of energetic ("hard") neutrinos can 
significantly increase when the WIMP mass becomes larger than 
the respective energy threshold for the production of heavy secondaries. 

\begin{figure}[!t]
  \centering
\includegraphics[width=3.5in]{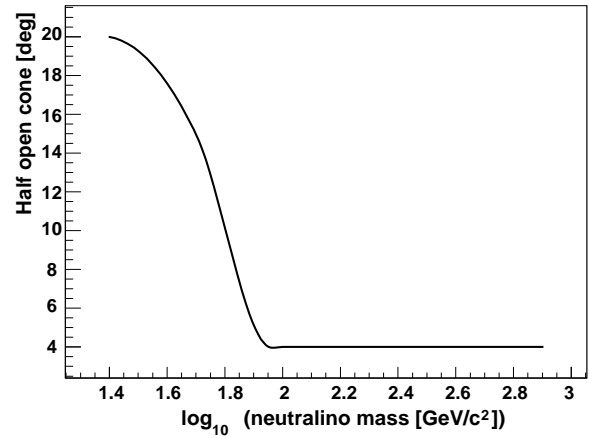}
\caption{Half cones for muon direction toward the Sun where the signal from a WIMP of
a given mass could be detected with $90\%$ probability.}
\label{ConeDM}
\end{figure}

\begin{figure}[!t]
  \centering
\includegraphics[width=3.5in]{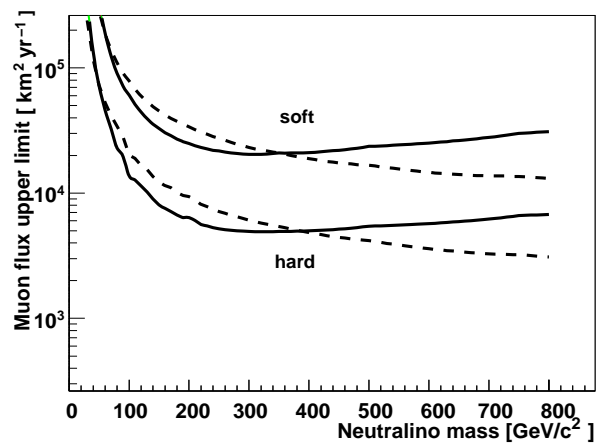}
\caption{The Lake Baikal NT200 upper limits at $90\%$ c.l. on the muon flux
from the Sun for a given WIMP mass (sample A - dashed line, sample B - solid line); 
"hard" neutrino fluxes are the lower two lines, "soft" the upper two lines.}
\label{UppSHMuon}
\end{figure}

From calculations of the expected number of muons $N_{\mu}$ according 
to (\ref{eqn : rate}), with the DM neutrino spectra from \cite{ref:Baksan96},
we estimated the size of the angular bin for which a WIMP signal could be
detected with $90\%$ probability. The found relation between half open
cone toward the Sun and WIMP mass is shown in Fig.\ref{ConeDM}.
The three angular bins in Table I are those for $90\%$ probability
of signal detection (M$\geq $90\,GeV, M=50\,GeV and M=30\,GeV, respectively). 
Further the upper limits on muon numbers were scaled
to WIMP masses. Table I gives the number of events (data, expected background and $90\%$
c.l. upper limit). We assumed neutrinos to be produced in two annihilation channels, 
b anti-b (soft channel) and $W^+W^-$ (hard channel) which define the respective
maximum energy $E_{up}$ for neutrinos. The resulting upper limits on muon 
fluxes at $90\%$ c.l. are shown in Fig.\ref{UppSHMuon}, as function of the WIMP mass, 
for the two selected samples and for soft and hard neutrino spectra
(normalized to $E_{thr}$=1 GeV). 
The upper limit on the WIMP-induced neutrino flux from the Sun
is found to be 
$F_{\nu}= {4.46}\times{10}^{10}km^{-2}\times{yr}^{-1}$
for $m_{WIMP}=100 GeV/c^2$ and for the annihilation channel $W^-W^+$ 
(sample B).

\section{Discussion}
Up to now all operating neutrino telescopes (see the
talks of the IceCube and ANTARES collaborations at this conference) and the past generation 
(Baksan \cite{ref:Baksan96},\cite{ref:Baksan2006} MACRO \cite{ref:MACRO}, Super-Kamiokande \cite{ref:SuperK}) 
reported no excess of upward going muons from the direction of the Sun when compared to the expectation for 
atmospheric neutrinos. This constrains both the annihilation rate in the Sun and (in concert with direct WIMP searches) 
the WIMP interaction cross section.

\begin{figure}[!t]
  \centering
\includegraphics[width=3.2in]{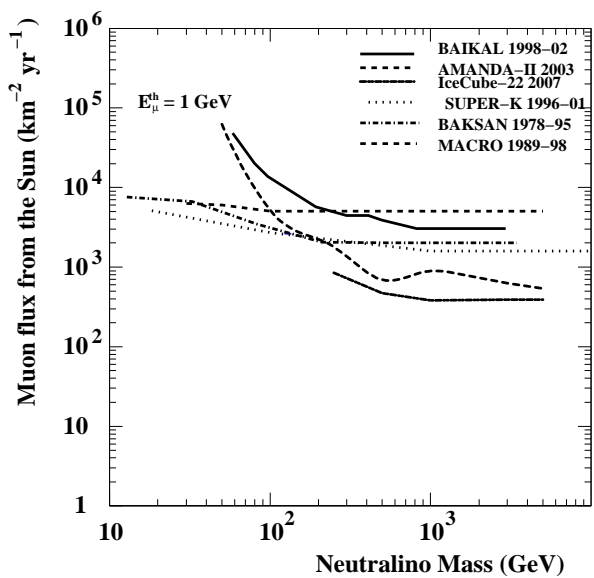}
\caption{Upper limits at $90\%$ c.l. on the muon flux from the Sun versus WIMP mass: 
Baikal NT200, MACRO \cite{ref:MACRO}, Baksan \cite{ref:Baksan96}, Super-Kamiokande \cite{ref:SuperK},
AMANDA-II \cite{ref:Amanda08} and IceCube 22-strings \cite{ref:IceQ09}.}
\label{baikal2_4}
\end{figure}

Upper limits on muon fluxes from the Sun obtained by neutrino telescopes are shown in Fig.\ref{baikal2_4}
(adapted from \cite{ref:Amanda08}; all limits are for 1 GeV threshold). 
The Baikal curve corresponds to hard neutrino spectra and 
is for $m_{WIMP}< 370 GeV/c^2$ obtained from sample B, 
above this value from sample A
(extrapolated up to 3\,TeV).
The AMANDA-II \cite{ref:Amanda08} and IceCube \cite{ref:IceQ09} limits are shown for hard neutrino spectra.

The presented results are preliminary, and allow to estimate 
the NT200 sensitivity for high energy neutrinos from DM annihilation 
processes in the Sun.

\section{Acknowledgments}


{\small
This work was supported in part by the Russian Ministry of Education
and Science, by the German Ministry of Education and Research,
by the Russian Found for Basic Research (grants 08-02-00432-a,
07-02-00791, 08-02-00198, 09-02-10001-k, 09-02-00623-a), by the grant of
the President of Russia NSh-321.2008-2 and
by the programm "Development of Scientific Potential in Higher Schools"
(projects 2.2.1.1/1483, 2.1.1/1539, 2.2.1.1/5901).
}

\end{document}